# Assessing the Dynamics of the Coffee Value Chain in Davao del Sur: An Agent-Based Modeling Approach

**Lucia Stephanie B. Sibala[1]\*, Giovanna Fae R. Oguis[2], and Novy Aila B. Rivas[3]**

[1]Department of Mathematics, Physics and Computer Science, University of the Philippines Mindanao, Davao City, 8000 Philippines

\*Corresponding Author: groguis@up.edu.ph

ABSTRACT

The study investigates the coffee value chain dynamics in Davao del Sur using an agent-based model. Three main factors driving interactions among key players were identified: trust, risk, and transaction costs. The model was constructed using NetLogo 6.3.0, and data from a survey questionnaire collected three data points from BACOFA members. Five cases were explored, with each scenario simulated 1000 times. Findings suggest that producers often sell to the market rather than the cooperative due to higher prices. However, producers tend to prioritize trust in buyers and their risk attitude, leading to increased sales to the cooperative. The producer's risk attitude significantly influences their decision-making, affecting performance outcomes such as loans, demand, and price changes. All three factors play a role and exert varying impacts on the value chain. So, the stakeholders' decisions on prioritizing factors in improving relationships depend on their priorities. Nonetheless, simulations show that establishing a harmonious system benefiting all parties is possible. However, achieving this requires adjustments to demand, pricing, trust, and risk attitudes of key players, which may not align with the preferences of some parties in reality.

INTRODUCTION

Coffee requires certain conditions to be produced, such as temperature and elevation above sea level (Scott, 2015). The Philippines, being located within the "Coffee Belt" geographical zone, can satisfy these conditions, and currently produces four varieties of coffee. Robusta is the most widely produced variety in the country, followed by Arabica, Liberica, and Excelsa. Historically, the Philippines used to be the world's fourth-largest exporter of coffee back in the 18th century (Uy, 2022). However, the *Philippine Coffee Industry Roadmap for 2021-2025* (2022) reveals a decline in coffee production in past the decade, leading to increased reliance on imports to meet the growing local demand and consumption. The decline of coffee production in the country is attributed to various factors including pests, diseases, and farmers shifting to other crops due to high maintenance costs (Department of Agriculture, 2022).

In recent years, producers have shifted to other crops due to high upkeep costs that outweigh the profit, despite the increase in coffee shops and buyers (Department of Agriculture, 2022). The production of coffee involves multiple factors, including limited bargaining power, market access, credit availability, and the quality of inputs, which entail the involvement of various players (Department of Agriculture, 2022; Sabroso and Tamayo, 2022). These factors emphasize that the challenges in the value chain are affected by the decisions made by individual players. Many dilemmas arise throughout the value chain among various stakeholders. Thus, this study aims to focus on the behavior of the key players in the value chain and how the changes in these behaviors affect the dynamics of the chain.

The coffee value chain encompasses various functions, including input supply provision, coffee bean production, primary and secondary processing, trading, marketing, retailing, and exports. Input supplies are sourced from the government, private nurseries, or fellow farmers in the form of investments, planting materials, and equipment. Generally, to negotiate prices more effectively and access larger markets, producers join cooperatives or form clusters to consolidate their coffee (University of Wisconsin-Madison, n.d.). Some producer cooperatives also buy production inputs together to secure better prices. Cooperatives possess machinery for primary and secondary processing. Though some producers have the means to process harvested cherries, cooperatives provide discounted access to machinery for their members, while non-members pay full price (Paramount Coffee, n.d.). Both fresh cherries and GCBs are then sold to local processors, cooperatives, large coffee companies, or specialty coffee shops.

Ideally, cooperatives should absorb most of the volume from the member producers, buying both fresh cherries and GCBs. However, producers choose to sell their processed GCBs and fresh cherries /directly to traders, local roasters and institutional buyers who has secondary processing and primary processing capabilities, respectively, to avoid the deduction of income when selling to cooperatives due to their loans from the organization. And usually, other markets offer high buying prices compared to the cooperative.

This study examines the coffee value chain of Davao del Sur, Mindanao, Philippines, as this province accounts for 44% of the total volume of coffee produced in the Davao region (*Coffee RSIP Davao Region*, 2019). It focuses on the key players and their behaviors within this province particularly in the value chain of Arabica coffee variety. In the global chain, the Arabica variety accounts for approximately 60% of the total coffee production and is recognized as a high-quality variety due to its flavor and aroma (MacDonnell, 2019; Perk Coffee, 2017). These coffee beans thrive in higher elevations and require a meticulous and rigorous production and processing approach. As a result, market prices for this variety are high.

The Arabica coffee value chain in Davao del Sur involves smallholder farmers, cooperatives with primary processing capabilities, and roasters that can also act as input suppliers, consolidators, and markets.

Several studies have aimed to enhance the country's coffee industry. For instance, Sabroso and Tamayo (2022) used a Data Envelopment Analysis (DEA) approach to measure the technical efficiency of coffee production in Davao City. The study identified factors contributing to efficiency and emphasized the need to prioritize and acknowledge these factors. Another study by Tan (2021) focuses on coffee processors in Amadeo and Silang, Cavite, using a Business-Model approach to examine their market positions and the impact of the growing demand-supply gap on the local value chain. Snowball sampling was used to identify and analyze the coffee processors and other chain actors involved in the study (Human Research Protection Program, 2010).

This study uses the same technique by Tan (2021) to identify the potential stakeholders in the chain. Moreover, an Agent-based modeling (ABM) approach based on Dijkxhoorn et al (2017) was used to examine the behavioral changes of the key players of the coffee value chain. This study aims to assess the dynamics of the coffee chain in Davao del Sur and provide insights that may contribute to the betterment of the chain and the coffee industry. ABM is a suitable tool that considers both performance outcomes and behavioral changes. Notably, a typical value chain has diverse players with distinct functions that pursue different interests and objectives. Oftentimes, these objectives may be conflicting (Dijkxhoorn et al., 2017). ABMs can effectively capture the resulting interactions and changes in the value chain that arise from these different objectives.

A research project by Dijkxhoorn et al. (2017) also utilized ABM as an assessment tool to evaluate changes in farm income, transaction costs, and behavioral patterns in value chain relationships within an intervention strategy program. They applied a Theory of Change (TOC) framework, with trust, risk, and transaction cost as key elements. To measure these elements, games were conducted involving participating and non-participating farmers in the program. The outcomes of these games were then used as inputs for their NetLogo model.

A Theory of Change (TOC) framework will be utilized to model these behavioral changes. In particular, the study will rely on  Dijkxhoorn et al.'s (2017) Theory of Change, which has been modified to fit the context of the study (Figure 1). This theory proves to be appropriate for the coffee value chain of Davao del Sur, supported by relevant literature on the actual and theoretical behavior of the stakeholders. The interactions among value chain agents in the model will be governed by trust, risk, and transaction costs (i.e., production, and post-production costs) while proxy measures will be used to quantify these factors.

The estimation of costs, production yields, and income values in the model will be based on data obtained from the Pre-test Survey Questionnaire conducted by the Value Chain Lab at UP Mindanao in August 2023 on Balutakay Coffee Farmers' Association (BACOFA) farmers. Data on stakeholders and market prices will be limited to secondary data and literature available up until 2023 and its previous years. Also, the model may adopt simplified assumptions, which might not fully capture the intricacies of the value chain due to these limitations. However, the simulation will serve as a proof of concept and establish a starting point for future ABM studies.

This study seeks to address the following research questions: (1) how significant are trust, risk, and transaction costs to the key players, particularly loan providers, producers, processors, and markets, within the value chain? (2) how do these key players respond to changes in these factors? And (3) how do these factors influence performance outcomes within the value chain, such as the cumulative number of loans in the system, the average yield production of all farmers, and the quantity of produce received by each market?

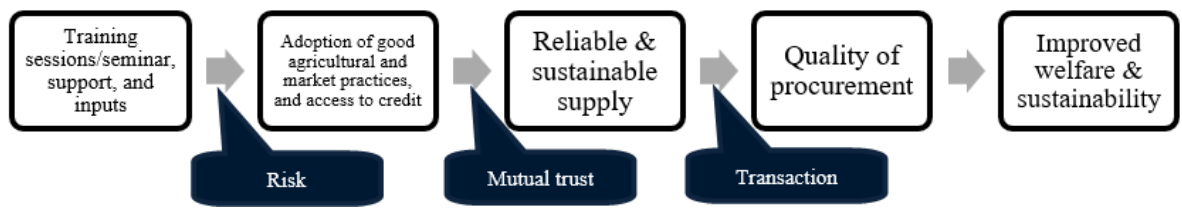

Fig. 1. Theory of Change Framework (Lifted from Dijkxhoorn et al., 2017)

Behavioral theories are frequently used as the theoretical framework in ABM applications to study smallholder farmers' adaptation behavior (Schrieks et al., 2021). These theories provide a basis for understanding and representing individual decision-making processes within the model. The study incorporates the framework proposed by Dijkxhoorn et al. (2017), a TOC approach with trust, risk, and transaction cost being the factors that influence the behaviors of the agents in the model.

TOC is a method that outlines how interventions are expected to bring about specific developmental changes. It is based on causal analysis using available evidence and involves the participation of relevant stakeholders in the coffee chain. The coffee industry follows a free market system with highly volatile prices for coffee beans (Dowding, 2017; IFPRI, 2020). Though most players act independently, trade dynamics between them can vary due to different factors such as trust, risk, and transaction costs.

Risk perception emphasizes the importance of considering individuals' risk preferences, as theoretical models and empirical studies have shown the value of incorporating risk into decision-making (Liu, 2013). On the other hand, mutual trust arises when individuals believe or expect positive and trustworthy intentions from others, even in uncertain situations (McAllister et al., 2017). Lastly, transactional costs refer to expenses influence by the nature of the exchange rather than the market price of goods or services (Robins, 1987).

MATERIALS AND METHODS

The flow of the methodology is presented in Figure 2. There are 4 steps involved: data collection and analysis, formulation of the ABM model, running simulations, and interpreting the results. Using snowballing technique, key players in the Arabica coffee chain were identified, including input suppliers, cooperatives, farmers, and markets. The behavior of these value chain actors was examined using relevant studies. Five cases were simulated with various parameter combinations. However, due to the limited data, model assumptions were incorporated to simulate the model. The model formulation will be further explained in the following sections.

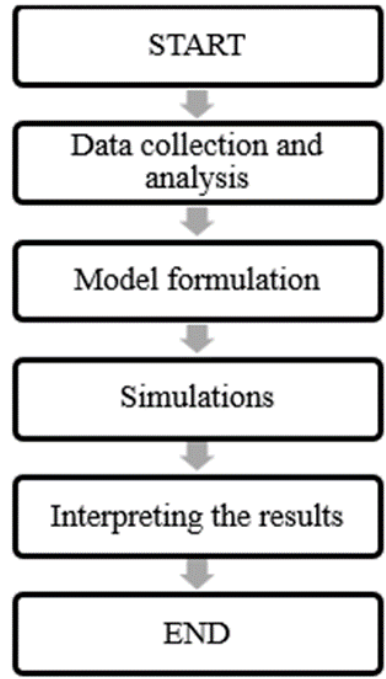

Fig. 2. Flowchart of Methodology.

The data used in the model were based on existing literature and secondary data to understand the behavior, interactions, and transactional costs of the value chain actors. This study used the data gathered by the Agri-aqua Value Chain Laboratory (AA VC Lab) during their pre-test survey in Balutakay, Bansalan, Davao del Sur coffee farmers in August 2023. Three (3) data points were used in running the preliminary analysis. The data provides insights on the farmer's practices in coffee farming, their motivations behind their choice of market for selling their coffee products, and their coffee yield, production and post-production costs, and overall income.

**Model Formulation**

The methodological framework of constructing the model follows a modified version of Ramanath and Gilbert's (2003) simulation research process, as shown in Figure 3. This framework provides a structured approach for designing and implementing the model, ensuring its coherence and reliability throughout the process.

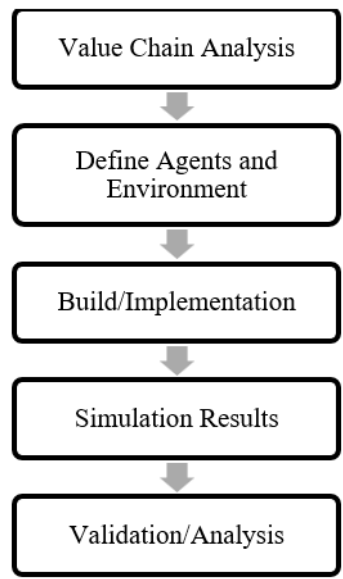

Fig. 3. Framework for ABM (Lifted from Ramanath and Gilbert, 2003).

Value chain analysis was conducted to understand the Arabica coffee chain in Davao del Sur. Figure 4 displays the Arabica coffee value chain in Davao del Sur. Moreover, Figure 5 illustrates the specific players of the chain that were identified using snowballing technique. There are two big cooperatives in Balutakay that involve coffee farming and production, the Balutakay Coffee Farmers Association (BACOFA) and Bagobo Tagabawa Farmers Managa Associations (BaTaFaMa). According to the Department of Trade and Industry (n.d) there are approximately 150 coffee farmers in Balutakay, Bansalan where around 68% are members of BACOFA and the remaining are members of BaTaFaMa (Maches, 2022; Sinag Coffee, 2020). Hence, in the simulation, we assumed that there are 200 smallholders' farmers produces Arabica coffee variety in Davao del Sur. Five markets were considered in the model based on the CRS dataset (2015) and Coffee RSIP Davao Region (2019), including Equilibrium, Monk's Blend, Coffee for Peace, Mt. Apo Coffee, and Paramount Coffee.

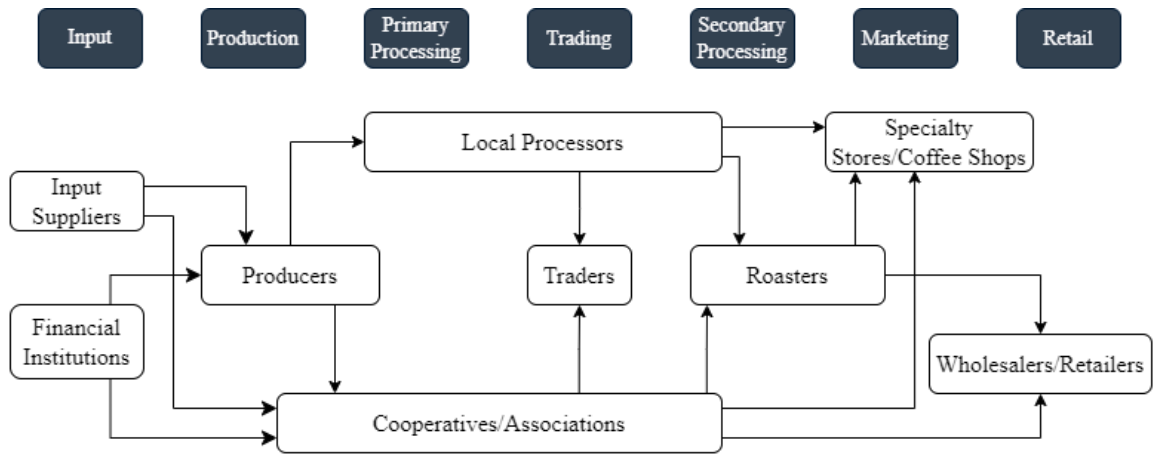

Fig. 4. Arabica coffee value chain in Davao del Sur.

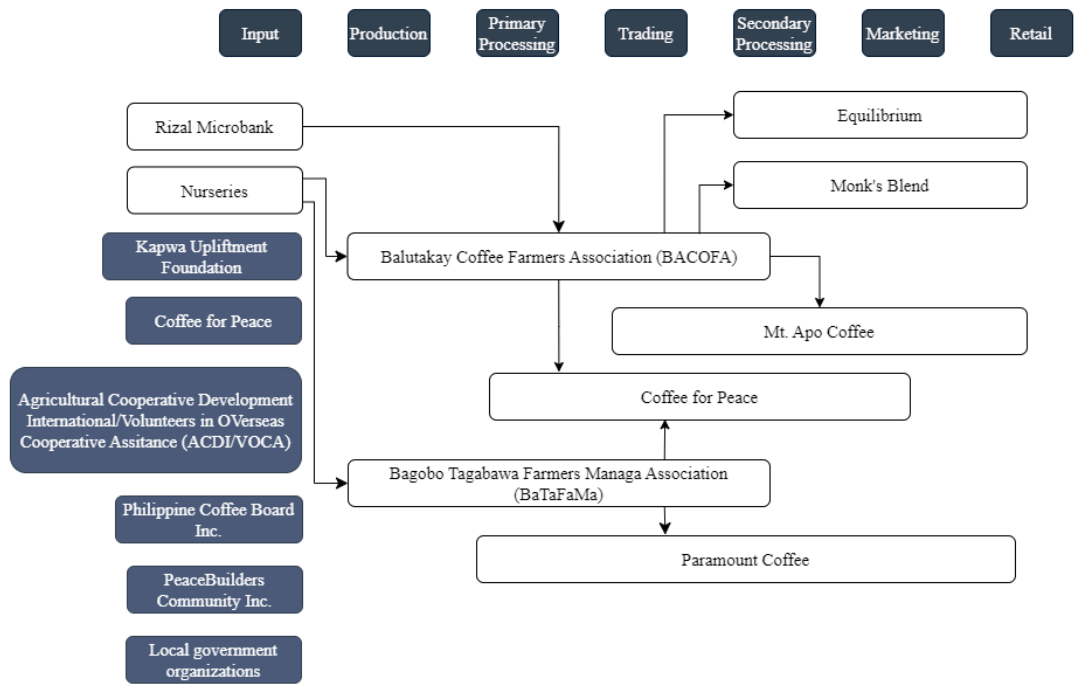

Fig. 5. Key players in the Arabica coffee value chain in Davao del Sur.

**Agent-based Model (ABM)**

Agent-based modeling approach was used to understand the behaviors of the agents in coffee production and marketing. These agents made the decisions based on their environment and interactions within the system. The developed model was bounded by several assumptions. These assumptions establish the conditions for valid inferences derived from the model, simplify the model, and did not capture the full complexity of the value chain. The following are the assumptions used in the model.

1. The producers only produce the Arabica coffee variety.
2. Producers will allocate the loans they acquire only for coffee production and processing. They will also prioritize repaying these loans.
3. The produced coffee consistently meets market quality standards.
4. The cooperatives and producers will transact with the markets directly.
5. Upstream players generate income or acquire capital through loans, without external financial support during simulations.
6. Producers will approach cooperatives that own the machinery to process fresh cherries into GCBs.
7. There will be no shortage of input supplies and finances for loans.
8. The model will not consider independent producers with primary processing capabilities.

*Corresponding Author: groguis@up.edu.ph

9. The independent producers will not choose to become members of the cooperatives.

10. Wholesale buying of inputs among groups will not be considered.

11. The players' mortality or the involvement of new players is not considered in the model.

12. The model will not consider the decline in coffee production due to aging trees.

13. Factors such as environmental, physiological, and some socioeconomic aspects have negligible influence on coffee production.

14. Factors other than risks, trust, and transactional costs have negligible influence on the decision-making process of the agents.

**Evaluation based on trust, risk, and transaction costs**

Interactions in the simulations were determined by the associated trust, risk, and transaction cost between involved agents (Equation 1).

Assuming that $x$ is an agent that provides service or resources, and $y$ is the recipient. For $x$ to provide $y$ with the service, $x$'s evaluation for $y$ must be greater than or equal the set evaluation threshold $T$. So, let $E_{xy}$ be $x$'s evaluation for $y$, then

$$E_{xy} = \omega_1 E_{xy}^t + \omega_2 \left( 1 - E_{xy}^r \right) + \omega_3 E_{xy}^c \qquad [1]$$

where $E_{xy}^t$, $E_{xy}^r$, $E_{xy}^c$ are the evaluation metric for trust, risk, and transactional costs, respectively, and $\omega_n$ for $n = 1, 2, 3$ is the weights associated with each metric.

The evaluation metrics represents proxy measures to understand the interaction and relationship between $x$ and $y$. Moreover, a beta distribution based on the model proposed by Wang and Singh (2010) was adopted in this study for measuring trust and uncertainty. Standard deviation of returns was used to estimate the volatility based on the study by Daly (2008), and the value function of the prospect theory by Kahnemann and Tversky's was used for measuring the evaluations based on costs (Wang et al., 2020).

*Corresponding Author: groguis@up.edu.ph

Weights in the different decision factors was assigned in the model similar to Kopp and Salecker's (2020) study. The weights $\omega_n$ ($n = 1, 2, 3$) indicate $x$'s priorities. These weights ranges between 0 and 1, and sum up to 1, ensuring that the evaluation metric $E_{xy}$ also falls within the interval of $[0, 1]$. These weights are based on empirical data and literature, such as the behaviors and motivations gathered by the CRS 2015 project. For instance, the CRS 2015 project data revealed that framers' main motivation for choosing markets was high prices, indicating that the cost evaluation function $E_{xy}^c$ will carry more weight in a farmer $x$'s evaluation of a market $y$. Literature suggests that these factors are also influenced by different aspects, including roles, interactions, price volatility, gains, and loss (Department of Agriculture, 2022; French et al., 1987; Mittendorf et al., 2019).

In particular, $x$'s trust evaluation for $y$ is

$$E_{xy}^t = \omega_4 C_y + \omega_5 I_{xy}^t + \omega_6 R_y^t \qquad [2]$$

where, $C_y$ is the value for cooperative membership, $I_{xy}^t$ is the value for trust based on interactions between x and y, $R_y^t$ is the value for trust based on y's reputations, and $\omega_n$ for $n = 4, 5, 6$ is the weights associated with each factor.

*Corresponding Author: groguis@up.edu.ph

Likewise, the weights $\omega_n$ $(n = 4, 5, 6)$ correspond to each function in the evaluation metric for trust $E_{xy}^t$. These new sets of weights pertain to priorities for the factors related to trust and may differ for each agent. For example, an agent $x$ may prioritize personal interactions with another agent $y$ rather than $y$'s reputation. The $C_y$, $I_{xy}$, $and$ $R_y$ are functions for cooperative membership, the resulting trust based on the interactions between $x$ and $y$, and the expected trust based on $y$'s reputation. These weights and functions also fall within the interval $[0, 1]$, and hence so does $E_{xy}^t$. The functions are given as

$$C_y = \{1, \quad if\ y\ is\ a\ cooperative\ member \quad 0, \quad if\ y\ is\ not\ a\ cooperative\ member \ , \qquad [3]$$

$$I_{xy}^t = \frac{\alpha_{xy}}{\alpha_{xy} + \beta_{xy}}, \qquad [4]$$

$$R_y^t = \frac{\alpha_y}{\alpha_y + \beta_y} \qquad [5]$$

where $\alpha_{xy}$ and $\beta_{xy}$ are the positive and negative interactions between x and y, respectively, while $\alpha_y$ and $\beta_y$ are the $y$'s positive and negative interactions with all agents, respectively. These values are updated after each new interaction.

On the other hand, $x$'s risk evaluation for $y$ consider the different areas of uncertainty. The evaluation metric for risk is given by

$$E_{xy}^r = \omega_7 I_{xy}^r + \omega_8 R_y^r + \omega_9 P_{xy} \qquad [6]$$

where $I^r_{xy}$ is the value for risk based on interactions between $x$ and $y$, $R^r_y$ is the value for risk based on $y$'s reputation, $P_{xy}$ is the value for risk based on $y$'s payments, and $\omega_n\ for\ n\ =\ 7, 8, 9$ are the weights associated with each factor.

The functions $I^r_{xy}$ and $R^r_{xy}$ compute for the variance $\sigma^2$ of the Beta distribution based on Wang and Singh (2010) given the same parameters used in $I^t_{xy}$ and $R^t_{xy}$. On the other hand, $P_{xy}$ pertains to the standard deviation $\sigma$ of the returns of $y$'s payments for $x$'s services based on the study of Daly (2008). Similar to the previous weights, the parameters $\omega_n\ (n\ =\ 7, 8, 9)$ are unique to the evaluation metric for risk and also depend on the agent's priorities for assessing risk. These functions are given in equations 7-9.

$$I^r_{xy}\ =\ \frac{\alpha_{xy}\beta_{xy}}{\left(\alpha_{xy}+\beta_{xy}\right)^2(\alpha_{xy}+\beta_{xy}+1)} \qquad [7]$$

$$R^r_y\ =\ \frac{\alpha_y\beta_y}{\left(\alpha_y+\beta_y\right)^2(\alpha_y+\beta_y+1)} \qquad [8]$$

$$P_{xy}\ =\ \sqrt{\frac{\sum\limits_{1}^{N}\left(r_i-\bar{r}\right)^2}{N-1}} \qquad [9]$$

where $r_i$, $\bar{r}$ and $N$ are the return of investment or asset i, mean of the return of asset i indicating a gain or loss, and N is the number of observations, respectively. Since $x$ lacks a reference point for comparison during the first payment and interaction with $y$, then $N > 1$. This function won't apply during these first transactions, hence the initial value of $P_{xy}$ is set to be 1, based on McAllister et al.'s (2017) theory regarding an individual's initial uncertainty and trust. Consequently, $x$'s cost evaluation is based on their losses and gains, possibly potential, during their interaction with $y$. This function is given as:

$$E_{xy}^c = \{ \left( \frac{z_{xy}}{z_r} - 1 \right)^{\omega_{10}}, \quad z_{xy} \geq 0 \quad 1 - \left( -\theta \left( \frac{z_{xy}}{z_r} - 1 \right)^{\omega_{11}} \right), \quad z_{xy} < 0 \qquad [10]$$

where, $z_{xy}$ is the amount of x owns after interacting with y, $z_r$ is the $x$'s reference point, $\omega_n$ $(n = 10, 11)$ are the weights associated with each factor, and $\theta$ is the loss aversion parameter.

In this scenario, $z_{xy}$ is the amount $x$ has after their interaction with $y$ while $z_r$ may pertain to the amount they used to provide $y$ or their money in hand, depending on the interacting agents. Then if $\frac{z_{xy}}{z_r} \geq 0$, then $x$ perceive it as a gain, otherwise it is a loss. With regards to losses, $x$'s evaluation is based on how large their loss is rather than just losing profit. Moreover, $\omega_{10}$ and $\omega_{11}$ corresponds to the degree of diminishing sensitivity towards gains and losses, as discussed by Wang et al. (2020). These parameters describe how sensitive $x$ is with each gain or loss. Each weight also falls between $[0, 1]$ and, unlike the set of weights introduced before, these is not sum up to 1. The loss aversion parameter $\theta$ describes how $x$ values a loss compared to a gain. To ensure that the function produces a value between $[0, 1]$, then it is also assumed that:

$$E_{xy}^c = \{ 1, \quad E_{xy}^c \geq 1 \quad 0, \quad E_{xy}^c \leq 0 \quad E_{xy}^c, \quad otherwise \qquad [11]$$

Once $x$'s evaluation of the costs reaches a certain threshold, all subsequent points will be considered equal, specifically either $1$ or $0$. In the gain function, the evaluation metric equals $1$ if $z_{xy} \geq 2z_r$, while it equals $0$ when $z_{xy} = z_r$. This is applicable because an interaction that results in $x$ gaining twice the amount they spent or own should have a significant impact on the evaluation. On the other hand, receiving amounts similar to what they spent or own will be considered less valuable by $x$. For the loss function, the outcome depends on the set parameters. Once the value of $\left(-\theta\left(\frac{z_{xy}}{z_r} - 1\right)^{\omega_{11}}\right) \geq 1$, then the function $E_{xy}^c = 0$. This means that agent $x$ would experience a significant loss and contribute nothing positive to the evaluation function.

*Corresponding Author: groguis@up.edu.ph

Table 1. Initial positive interactions between agents $x$ and $y$.

| $x$ \ $y$ | Cooperative | Farmers | Loan Providers | Markets |
|---|---|---|---|---|
| **Cooperative** | No interaction between agents. | Contribute to the needed supply. | No interaction between agents. | Pay the cooperative. |
| **Farmers** | Pay the farmers. | No interaction between agents. | Provide access to credit. | Pay the farmers. |
| **Loan Providers** | No interaction between agents. | Repay their loans. | No interaction between agents. | No interaction between agents. |

*Corresponding Author: groguis@up.edu.ph

| | Cooperative | Farmers | Loan Providers | Markets |
|---|---|---|---|---|
| $y$ | | | | |
| $x$ | | | | |
| **Markets** | Meet market demand. | Receive produce from farmers. | No interaction between agents. | No interaction between agents. |

## Agents and their attributes

Each type of agent will follow a set of rules that mirror real-life agents in the value chain. While most agents will operate independently, their decisions to engage in trade with one another will depend on the evaluation metric. The evaluation process will differ for each role. Table 2 summarizes the initial positive interactions from $x$'s perspective. The first row lists the providing agents $x$ while the column headers list the receiving agents $y$ Moreover, Figure 6 presents the proposed discussion in this section, highlighting what agents consider positive or negative interactions.

**Input suppliers.** There will be two types of input suppliers: loan providers and a nursery. The nursery will act as a passive agent, offering necessary inputs, and will be accessible to all producers. Meanwhile, loan providers will provide finances to producers that seek loans, the success of acquiring these loans will be based on their evaluation of the producers. Positive interactions for this agent refer to producers' timely repayment of their loans, while failure to do so will be a negative interaction. If the producer fails to acquire the necessary loans, they will proceed with managing their farms, which could potentially result in reduced yields.

*Corresponding Author: groguis@up.edu.ph

**Cooperatives.** The model will include two cooperative agents that will be connected to all their members. These agents will gather the fresh cherries and GCBs produced by cooperative members for collective marketing.

When a farmer consistently contributes a certain amount within a given time step, it will be seen as positive evidence of their relationship with the cooperative, and their reputation as a smallholder. Conversely, not contributing anything will be considered negative evidence. Both cooperative members and non-members will be required to pay post-production fees to the cooperatives. In terms of cooperatives and markets, receiving a fair price and avoiding lowball offers would ideally be seen as positive evidence by the cooperative towards a market. However, this will depend on the availability of market price data. To simplify matters, receiving payment from the market will be considered as positive evidence in this context.

**Farmers.** Smallholder farmers will produce and process coffee beans. Farmers will either be a cooperative member or an independent. As a result, the cooperative function $C_y$ will not influence the evaluation of the farmers. Moreover, the evaluation of risk based on the payments from their respective buyers $P_{xy}$, will be specific to each individual buyer. Based on the 2015 CRS project data, producers will prioritize selling to buyers that offer greater prices. According to the same data, production costs were Php 25.40 per tree while post-production was Php 11.44 per tree. These trees produced an average of 3.24kg of fresh cherries or 0.58kg of GCBs per tree. Hence, the post-production per 3.24kg of fresh cherries or 0.58kg of GCBs in the simulation will be Php 25.40 and Php 11.44.

*Corresponding Author: groguis@up.edu.ph

Farmers will calculate the needed inputs based on the number of trees they own. Then, they will decide whether to loan or not since many prefer not to, due to the fear of not being able to repay them (Sabroso and Tamayo, 2022). If the loan provider refuses to give them loans, then the producer will consider this as negative evidence. As mentioned, farmers that fail to provide inputs for their farms will result in lower yields of coffee cherries. After harvesting fresh cherries, farmers will make decisions on whether to sell them as fresh cherries or perform primary processing which includes drying, dehulling, washing, and sorting the produce. They will also determine where to sell based on the evaluation metric. If none of the markets meet the evaluation metric, the farmer will be compelled to sell their cherries at a lower price. Nonetheless, each payment towards the producer will be regarded as positive evidence. Farmers will process as much as they can afford and sell the rest as coffee cherries. By evaluating the existing markets, farmers will decide to whom they will sell. Then, at the end of each time step, they will decide to pay their loans if they have the finances to do so.

**Markets.** The market agents that represent Equilibrium and Coffee for Peace will buy GCBs with prices and demands based on the *Coffee RSIP Davao Region* (2019). On average, GCBs are bought for Php 220.00 while fresh cherries are bought for Php 35.00 (Department of Agriculture, 2022). Equilibrium will buy GCB for Php 250.00 to Php 300.00. Coffee for Peace will buy GCBs for Php 180.00 to Php 300.00, and fresh cherries for Php 38.00 to Php 40.00. Other markets will set a price based on PSA's data on farmgate prices for GCBs and fresh cherries and CRS farmers' selling prices.

*Corresponding Author: groguis@up.edu.ph

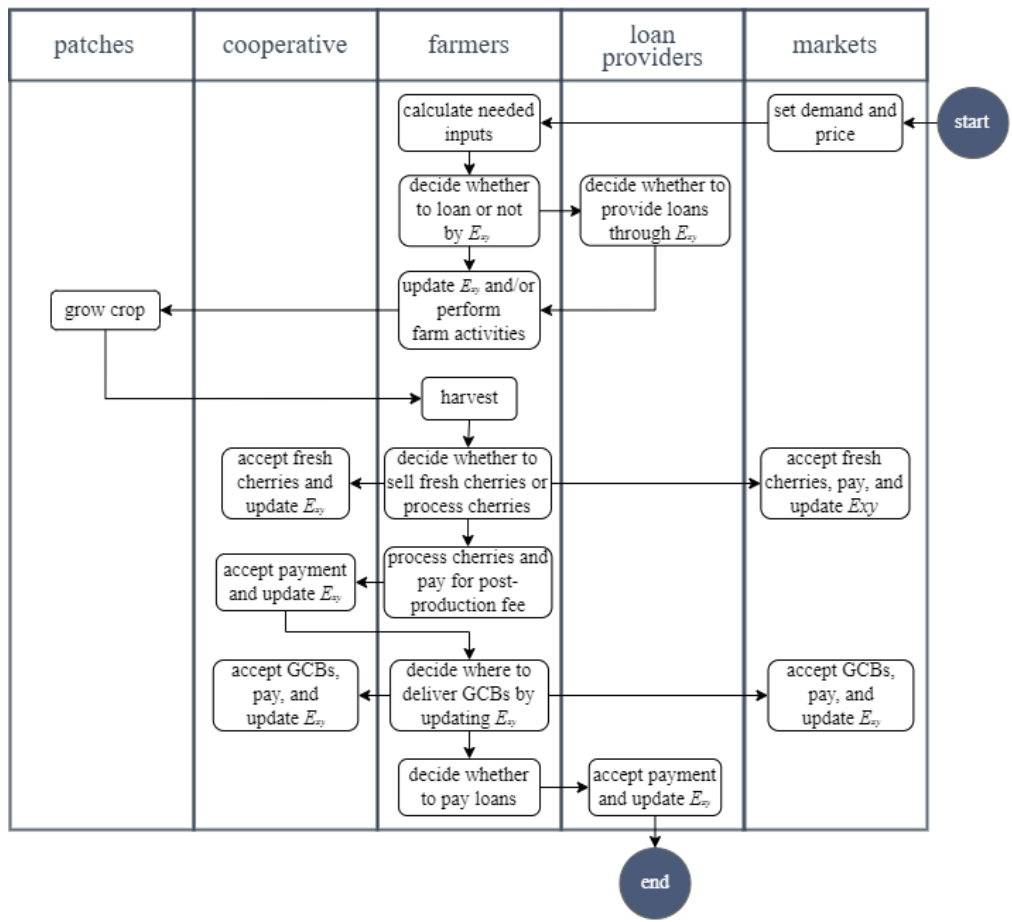

Fig. 6. Overview of the process during a time step in the NetLogo model.

Markets will have a demand per time step and expect their cooperative suppliers to meet a certain amount of this demand. The produce of their main suppliers will be prioritized. However, failure to meet this requirement will be treated as negative evidence and will affect their relationship. It is important to note that market demands are equivalent to their purchasing capacity, and they will not accept more than their set quantity.

**Modeling environment**

*Corresponding Author: groguis@up.edu.ph

The active entities will be randomly distributed in the modeling environment, with cooperative members located near their respective cooperative agents. Each farmer will own a patch of land equivalent to 1-3 hectares. The color of the link between agents will indicate whether their evaluation has surpassed a set threshold. A green link signifies that the agent's evaluation has met or exceeded the threshold, while a red link indicates that the evaluation falls below it.

The simulation will progress in discrete yearly increments, with each time step representing one year. The simulation will cover a span of 20 years. To ensure statistically reliable results, the study will follow the methods outlined by Abbi and Peters' (2018) and the guidance provided by Helbing (2012). Performance outcomes will be assessed through graphs showing the mean and standard deviation, considering noise and convergence. If these values show convergence or stability, it indicates that the simulation duration is sufficient to evaluate patterns and dynamics within the value chain, while considering computational costs and time constraints. However, if the system reaches a stable state within a shorter timeframe or the patterns and dynamics are inconclusive, the study will consider adjusting the simulation duration accordingly.

**Model parameters**

The model's default parameters, which relate to the environment and real-world entities, are summarized in Table 3 and Table 4. These parameters will serve as the initial conditions moving forward. Fixed parameters cannot be modified through the user interface in NetLogo, while variable parameters can be adjusted by the user. The variables for farms, trees, and demand will have discrete values, while the remaining variables will be continuous but rounded to two decimal places. These values are derived from secondary data and existing literature.

Table 3. Variable default model parameters.

| Parameter | Value | Unit | Source |
|---|---|---|---|
| CFP-demand | 4 | tons/year (GCB) | *Coffee RSIP Davao Region* (2019) |
| Equilibrium-demand | 10 | tons/year (GCB) | *Coffee RSIP Davao Region* (2019) |
| Other-demand | 7 | tons/year (GCB) | Estimated |

Table 4. Fixed default model parameters.

| Parameter | Value | Unit | Source |
|---|---|---|---|
| Farms | 1-3 | ha/farmer | *Coffee RSIP Davao Region* (2019) |
| Trees | 693 | trees/ha | Department of Agriculture (2022) |
| Basic-expected-yield | 3.24 | kg/tree (fresh) | CRS 2015 Project |
| Prod-cost | 25.40 | Php/tree | CRS 2015 Project |
| Post-harvest-cost | 11.44 | Php/tree | CRS 2015 Project |

| Parameter | Value | Unit | Source |
|-----------|-------|------|--------|
| GCB-farmgate-price | 102.61-211.11 | Php/kg | PSA (2023) |
| FC-farmgate-price | 35.00-40.00 | Php/kg | CRS 2015 Project |
| Equilibrium-price | 250.00-300.00 | Php/kg | *Coffee RSIP Davao Region (2019)* |
| CFP-FC-price | 38.00-40.00 | Php/kg | *Coffee RSIP Davao Region (2019)* |
| CFP-GCB-price | 180.00-300.00 | Php/kg | *Coffee RSIP Davao Region (2019)* |

These values may be adjusted depending on the findings of the value chain analysis. During the simulation, farmgate prices and land sizes for farmers will be randomly generated within predetermined ranges. These values in particular will be continuous.

**Evaluation function parameters**

As previously mentioned, the weights allocated to the factors of the evaluation function will fall between the interval of $[0, 1]$, and the weights for $E_{xy}$, $E_{xy}^t$, and $E_{xy}^r$ will sum up to $1$. If there is a lack of clear distinction between the priorities of agents for certain factors, then these weights will be assumed to be equal. Consequently, if there is a lack of evidence that the agent cares for some of these factors, then these weights will be $0$.

The initial values of the evaluation threshold $T$, and all the discussed weights $\omega_n \, (n = 1, ..., 11)$ will be based on a thorough analysis of the value chain. However, the weights $\omega_n$ wherein $\sum \omega_n = 1$ will be initially fixed and assumed as equal, indicating equal priority of these factors in each metric for all agents.

Additionally, the weights of the cost function, $\omega_{10}$ and $\omega_{11}$, will also be fixed to $0.5$. This ensures a moderate response of agents to both gains and losses. Notably, these set values may change during implementation. The loss aversion parameter $\theta$ will also be set to $2$, indicating that losses have twice the impact compared to gains of the same magnitude. The discussed values are summarized in Table 5.

On the other hand, other variables in the evaluation function such as $\alpha, \beta, r, p,$ and $z$ will depend on the interactions, price, and payments between agents during the simulation. Hence, these will vary at each time step. Non-monetary variables (such as $\alpha$ and $\beta$) will initialize with a value of $1$, indicating that agents will be trust-inclined during the first interactions. Meanwhile monetary-related variables will be set to the values mentioned in the previous section.

Table 5. Fixed default evaluation parameters.

| Parameter | Value |
|---|---|
| $\omega_4 = \omega_5 = \omega_6$ | $0.\overline{3}$ |

*Corresponding Author: groguis@up.edu.ph

| Parameter | Value |
|---|---|
| $\omega_7 = \omega_8 = \omega_9$ | $0.\overline{3}$ |
| $\omega_{10} = \omega_{11}$ | $0.5$ |
| $\theta$ | $2$ |

**Simulations**

The study will begin by simulating the default parameters and initial conditions. Then, simulations will be conducted for different parameter combinations. To ensure reliable results, $1000$ simulations will be performed for each combination, following the recommendation of Abbi and Peters (2018).

However, the actual number of simulations may vary depending on the time required per simulation and the analysis of results. If the number of simulations is insufficient or excessive for conclusive and representative findings, the number will be adjusted accordingly. The study will explore three scenarios for each weight $\omega_n$ ($n = 1, 2, 3$) wherein each weight is of high priority, equal priority, and not a priority. These scenarios will yield seven cases, as summarized in Table 6. Likewise, aside from its default value, three values will also be explored for the evaluation threshold $T$: $1$, $0.5$, and $0$. These values aim to observe scenarios where agents display strict, neutral, and lenient attitudes in evaluating other agents. This is summarized in Table 7.

Table 6. Scenarios for a weight $\omega_n$.

*Corresponding Author: groguis@up.edu.ph

| Scenario | Possible values | Cases |
|---|---|---|
| one weight is highly prioritized | $1, 0$ | 3 |
| all weights are equally prioritized | $0.\overline{3}$ | 1 |
| one weight is not prioritized | $0, 0.5$ | 3 |

Table 7. Scenarios for a threshold $T$.

| Scenario | Value |
|---|---|
| strict attitudes | 1 |
| neutral attitudes | $0.5$ |
| lenient attitudes | 0 |

The study will examine a total of $113$ parameter combinations, considering different agents, weight scenarios, and threshold scenarios in addition to the default parameters and initial conditions. Each combination will be evaluated using the default initial conditions and parameters. Simulations for each combination will be conducted $1000$ times. Throughout these simulations, the following graphs will be generated:

1. A graph of the number of loans per time step.

2. A graph of the number of fresh cherries and GCBs each market receives per time step.

3. A graph of the price each market sets per time step.

4. The average trust, risk, and transaction cost evaluations of each unique agent per time step.

*Corresponding Author: groguis@up.edu.ph

**Interpreting the Results**

The graphs will provide insights into the agents' behaviors and the dynamics of the value chain. These graphs will be analyzed to identify and document notable performance outcomes. A panel regression analysis will be conducted to compare these outcomes with the default parameters and initial conditions. The findings will be described qualitatively, presenting a descriptive discussion of the results. This discussion will cover observed patterns, behaviors, and speculated dynamics within the value chain.

**Panel regression**

A panel regression analysis will be conducted to further explore the relationships and effects of the parameter combinations on the outcomes of the simulation. The panel regression will consider the performance outcomes of different parameter combinations as dependent variables, and the time steps in the simulation as independent variables. The performance outcomes include the number of loans, average yield production, number of fresh cherries and GCBs received by each market, the price set by each market, and the average trust, risk, and transaction cost evaluations of each unique agent.

To ensure the validity of the model, the findings from the panel regression analysis will also be compared to both theoretical expectations and real-world behaviors of agents in the coffee industry. The *Coffee RSIP Davao Region* (2019) and relevant literature will serve as references for establishing the expected behavior of real-world agents. Any deviations observed in the simulation outcomes will be further investigated to understand the underlying factors and potential implications.

*Corresponding Author: groguis@up.edu.ph

## Software and Requirements

The ABM model will be developed using NetLogo, specifically on version 6.3.0. The research study will be conducted on a device equipped with an Intel(R) Core(TM) i3-1005G1 CPU operating at a clock speed of 1.20GHz, and a RAM capacity of 8.00 GB. The system running the model will be a 64-bit operating system, utilizing an x64-based processor architecture.

RESULTS

This section presents the outcomes generated by the ABM model and offers an in-depth analysis of the findings. Through a comprehensive examination of the results, this chapter aims to provide valuable insights into the dynamics and implications observed within the study.

## Proxy Measures

The proxy measures for trust, risk and transaction costs were simplified according to the data obtained from the Pre-test Survey Questionnaire.

Firstly, a producer's trust in the cooperative was influenced only by the cooperative's willingness to provide loans and the price it sets compared to the market. So, the producer's trust levels influence the market they choose to sell to and their risk attitude.

*Corresponding Author: groguis@up.edu.ph

On the other hand, the buyers' trust was measured simply according to the producer's ability to repay loans and supply their demand. The trust of a single producer is hence defined by

$$t\left(x_n\right) = \{0, \ x < 0 \ random(UT, LT), \ x = 0 \ t\left((x-1)_n\right) + \sum_{i=1}^{2} w_i u_{i'} \ x : \qquad [22]$$

where,

$t\left(x_n\right)$ : the trust of producer $n$ at time $x$

$UT$ : the upper limit for the initial trust of producers

$LT$ : the lower limit for the initial trust of producers,

$random(UT, LT)$ : a function that generates a random number between the limits

$w_i(i = 1, 2)$ : the weight associated with the producer's trust

$u_i(i = 1, 2)$ : the resulting effect of factor on the producers' trust

On the other hand, the trust that is portrayed in the graphs shown are defined by

$$T\left(x_N\right) = \frac{\sum_{n=0}^{N} t\left(x_n\right)}{N} \qquad [23]$$

where,

$T\left(x_N\right)$ : the average trust of all the producers at a time $x$

$N$ : the total number of producers

*Corresponding Author: groguis@up.edu.ph

Secondly, the risk-attitude is simply exclusive to producers. This attitude is influenced by the producer's trust, fluctuations in the market prices, their financial situation, and a comparison between the cooperative and the market's prices. The financial situation specifically refers to whether they have sufficient funds after loan repayments and the risk of not paying these loans. The producer's risk attitude influences their decision to sell to the market and their willingness to repay loans timely. Similarly, the risk-attitude of a single producer is defined by

$$r\left(x_n\right) = \{0,\ x < 0\ random(UR, LR),\ x = 0\ r\left((x-1)_n\right) + \sum_{i=3}^{8} w_i u_i,\ x \qquad [23]$$

where,

$r\left(x_n\right)$ : the risk-attitude of producer $n$ at time $x$

$UR$ : the upper limit for the initial risk of producers

$LR$ : the lower limit for the initial risk of producers

$w_i(i = 1, 2, ..., 8)$ : e weight associated with the producer's risk-attitude

$u_i(i = 1, 2, ..., 8)$ : the resulting effect of factor on the producers' risk-attitude

Similar to the trust of the producer, the risk-attitude that is portrayed in the graphs are defined by

$$R\left(x_N\right) = \frac{\sum_{n=0}^{N} r\left(x_n\right)}{N}$$

*Corresponding Author: groguis@up.edu.ph

where,

$R(x_N)$ − the average risk-attitude of all the producers at a time $x$.

Thirdly, the transaction cost is also primarily focused on the producers. A noteable result from the survey is that there isn't a significant difference in the preparation and costs involved between selling to the market and the cooperative, except for the buying price set by the buyer. Moreover, producers primarily decided to sell to these buyers based on their high buying price. In the model, transaction costs influence their capacity to repay loans, their risk attitude, and their choice of selling to their buyers. In the model, the producer's available funds is also defined by

$$c(x_n) = \{0, \ x < 0 \quad \alpha_n \beta_0^0 + rand(UC, LC), \ x = 0 \ c((x-1)_n) + \delta(\alpha \qquad [24]$$

where,

$c(x_n)$ : the producer $n$'s money at time $x$

$UC$ : the upper limit for off-farm and non-farm income of producers

$LC$ : the lower limit for off-farm and non-farm income of producers

$\alpha_n$ : the number of trees owned by producer $n$

$\delta$ : the producer's decision to sell to either the cooperative $(0)$ or the market $(1)$

$\beta_x^{\delta}$ : the price of coffee offered by buyer $\delta$ at time $x$

*Corresponding Author: groguis@up.edu.ph

$\gamma_x$ : the amount of loans the producer repays at time $x$

## Simulations

Five cases were explored. In the first case, all members sell only to the cooperative, mirroring the pre-test survey. In the second case, we introduce another buyer with assumed properties. The third scenario involves setting prices based on the supply and demand of each buyer. The fourth scenario changes the buyers' demands, while the fifth adjusts the weighting of trust, risk, and costs in the evaluation functions. For each case and varied changes, simulations were conducted 250 times to account for the stochastic parameters in the model. The simulations also stop at the 1000th tick, equivalent to 1000 months.

The model interface includes eight plots, each depicting key aspects. The upper left plot shows the producer's trust in the cooperative and market, reflecting risk attitudes' impact on buyer trust. The upper right plot displays the trust levels of both producers and buyers. The middle left plot indicates the system's loan count. In the middle center, the fourth plot represents the producer's trust in buyers. The middle right plot focuses on the producer's risk attitude. The lower left plot visualizes monthly coffee percentages sold to each buyer, while the lower middle plot illustrates buyer trust in producers. Lastly, the eighth plot reveals monthly prices set by the buyers.

*Corresponding Author: groguis@up.edu.ph

**Case 1**

In Case 1, where only one buyer is involved, market competition is absent. The majority of simulations led to the generation of very similar results across all simulations. That is, due to the lack of market competition, the producer's trust stabilizes at a very low value – around 0.2. On the other hand, the producer's risk attitude initially rises but eventually falls, hovering around the neighborhood of 0. The producer's trust in the cooperative also experiences a significant decline as their risk attitude increases. Additionally, the loans across all the simulations remain high around 150, often reaching lower values (about 120) when the producer's risk attitude increases and rise when their risk-attitude decreases.

**Case 2**

In Case 2, market competition is introduced, but significant variables like the price set by buyers remain randomized, leading to more interesting results. Across most simulations, the producer's trust in the cooperative and markets generally remains consistent. Furthermore, while loans are notably lower than in Case 1, they tend to converge toward a similar value. Since the prices for both market and cooperative remain random, the demand supplied also seems to fluctuate as much. However, we can see that the market is often supplied more than the cooperative.

**Case 3**

In Case 3, the buyers' prices are no longer random. Results show that it has similar findings to Case 2, with the added insight into how changes in price influence other key aspects of the system. The producer's trust in buyers and risk-attitude remains consistent across all cases. It can also be observed that a low risk-attitude results in the cooperative being given most of the supply. Otherwise, producers tend to sell 30% to the cooperative and 70% to the market, mirroring real-world scenarios. The number of loans in the system did not change as much, except for the significant increase, as shown in Figure 1, where producers' risk attitude is nearly zero and they start to sell exclusively to the cooperative.

*Corresponding Author: groguis@up.edu.ph

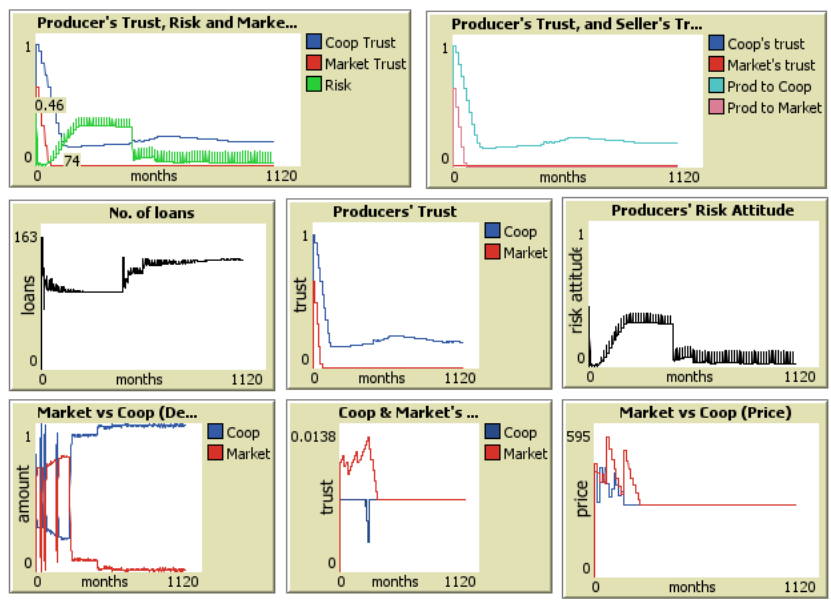

Fig. 7. Case 3 on NetLogo.

**Case 4**

In Case 4, we investigate eight scenarios involving incremental changes in demand at 10% intervals. The scenario where the cooperative demands 70% and the market demands 30% is omitted, as this value serves as the default and is already observable in Case 3.

*Corresponding Author: groguis@up.edu.ph

**Cooperative demands 10%, while market demands 90%.** Most simulations follow two trends: one where the risk attitude dips below the producer's trust towards the cooperative, and another where it is significantly higher. Nonetheless, all of these values eventually approach a certain number, around 0.2-0.4. However, in this scenario, producers opt to sell more to the market and less to the cooperative due to lower demand. Consequently, the market's trust declines, since the demand is high and the producers are unable to meet these requirements. In turn, market prices fluctuate greatly. Simultaneously, the cooperative's prices remain relatively stable due to their demand being satisfied.

**Cooperative demands 20%, while market demands 80%.** This scenario does not differ significantly from when the cooperative demands 10%, while the market deands 80%.

**Cooperative demands 30%, while market demands 70%.** At the start, there is increased activity in this scenario, with buyers adjusting their prices to meet demand. Additionally, there are more fluctuations observed in the number of loans. However, over time, both buyers converge to a similar price while still meeting their respective demand levels. It is noteworthy that as the risk attitude stabilizes, the perturbations stop. Also, producers with lower risk attitudes tend to have higher loan values.

**Cooperative demands 40%, while market demands 60%.** There is notably more activity at the start of the simulation when the producer's risk attitude is low. As the producer's risk attitude increases, the perturbations in buyers' demand tend to approach a 70-30 allocation faster. However, the levels of loans and the producer's trust in the cooperative and market remain consistent and similar to previous scenarios.

**Cooperative demands 50%, while market demands 50%.** The simulations have similar results to the previous one, however this case returned with a unique result. That is, instead of eventually approaching a value similar to the producer's trust to the cooperative, it resulted into a complete dip a few months in. Nonetheless, similar to a previous scenario in Case 3, a sudden decline in the producer's risk attitude led to an increased proportion of sales to the cooperative rather than the market. A lower risk attitude also affected the market's trust in the producer, as shown by its lower levels and greater fluctuations. Additionally, a lower risk attitude coincides with an increased number of loans.

**Cooperative demands 60%, while market demands 40%.** Once again, it is evident that a lower risk attitude corresponds to higher loan values and increased sales to the cooperative. In most simulations, a greater allocation is directed towards the market in comparison to the cooperative. Additionally, there is a correlation between fluctuations in buyers' demand and their trust in producers.

**Cooperative demands 70%, while market demands 30%.** This case is set as the default. Therefore, this scenario mirrors the simulations from Case 3.

**Cooperative demands 80%, while market demands 20%.** In this scenario, the producers often sell to the market. The price also fluctuates more and is due to the cooperative's unmet demands, leading to the market changing its prices in response. Nevertheless, the allocation eventually approaches 70-30 in favor of the market. Additionally, the prices stop fluctuating at a certain point in the simulation.

*Corresponding Author: groguis@up.edu.ph

**Cooperative demands 90%, while market demands 20%.** The final scenario is similar to the previous one. However, there are greater fluctuations in both demands met and buyers' prices. So, most simulations.

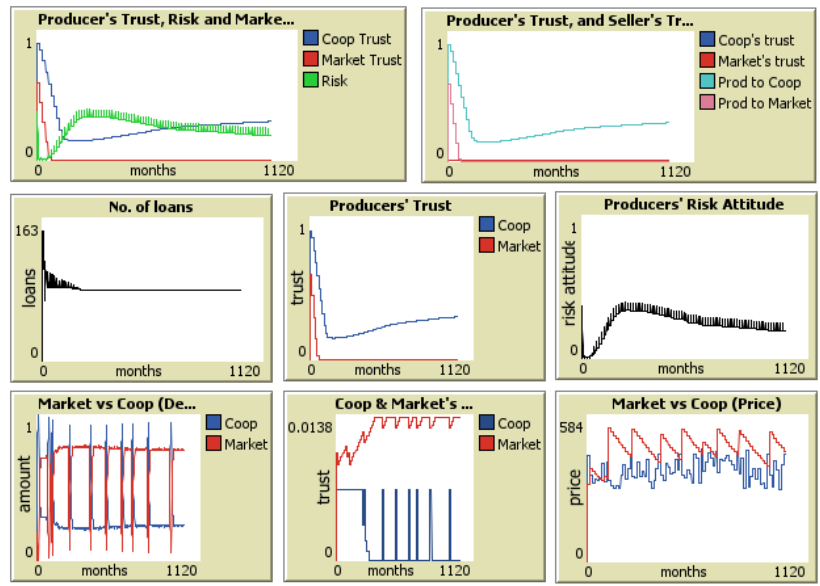

Fig. 8. Case 4, cooperative demands 90%, while market demands 10% on NetLogo.

**Case 5**

In Case 5, we investigate seven scenarios involving incremental changes in trust, risk, and transactional costs towards the evaluation function of the producer.

**33% trust, 33% risk, and 33% transactional costs.** The simulations result in an increased fluctuations in the producer's supplies for the cooperative and market. Despite that, there is still a similar pattern wherein a lower risk attitude leads to a larger portion of sales sold towards the cooperative. Conversely, a higher risk attitude resulted in increased sales in the market and, consequently, a lower number of loans. Notably, the simulations showed that there were more variations in the producer's trust in the cooperative. The buyers' prices show that the cooperative did not need to raise prices since its demands were adequately met. In contrast, the market initially set higher prices but eventually approached values similar to the cooperative while still meeting its demands.

*Corresponding Author: groguis@up.edu.ph

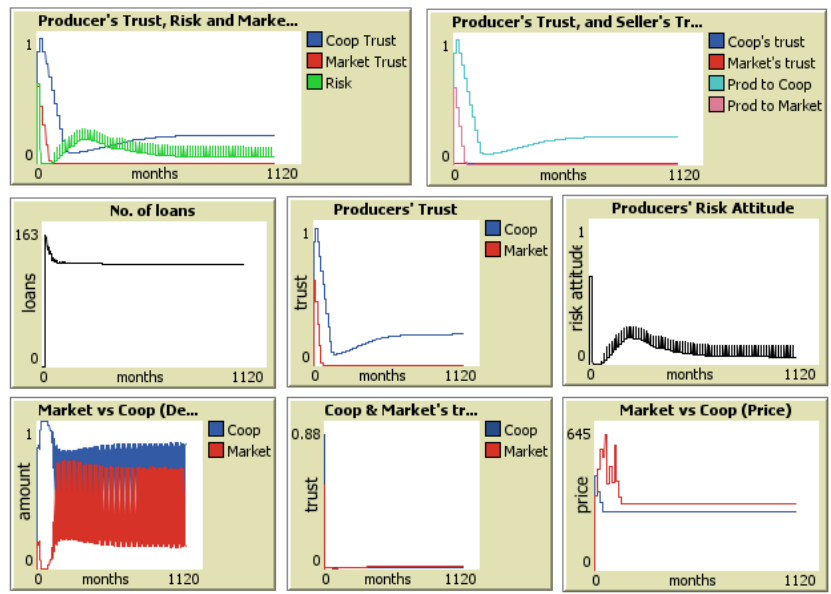

Fig. 9. Case 5, the producer prioritizes trust towards the buyer by 33%, the buyer's price by 33%, and their risk attitude by 33% in NetLogo

**100% trust.** In this scenario, the producer prioritizes trust towards buyer by 100%. Since the producer only prioritized trust towards the buyer, the majority of sales went toward the cooperative. All simulations consistently produced plots closely resembling Figure 4.

*Corresponding Author: groguis@up.edu.ph

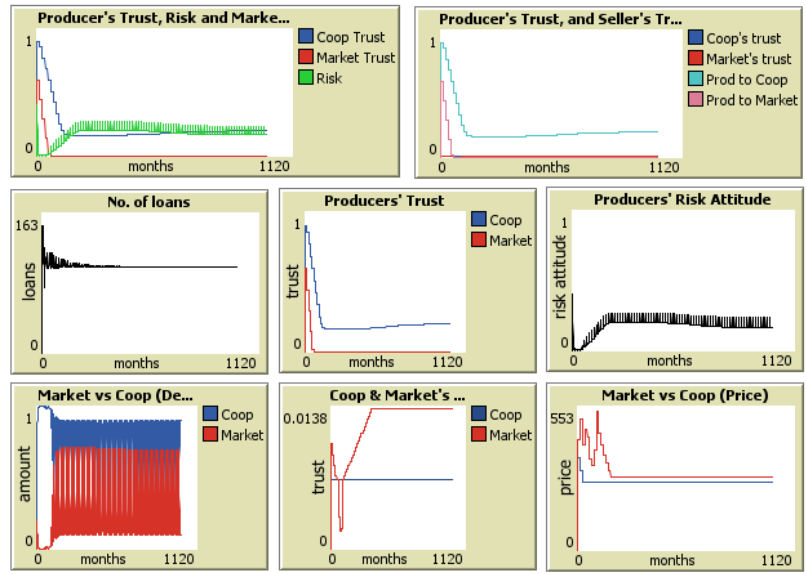

Fig. 10. Case 5, the producer prioritizes trust towards the buyer by 100% in NetLogo.

**100% transactional costs.** In this scenario, the producer prioritizes the buyer's price by 100%. This case is set as the default for all the previous cases, so this scenario mirrors the simulations from Case 3.

**100% risk-attitude.** In this scenario, the producer prioritizes their buyer's price by 100%. This scenario shares similarities with the producers prioritizing trust towards buyers, except that there are fewer fluctuations in demand, and most of the sales go to the cooperative. In most simulations, the producer's risk attitude tends to decline. Notably, regardless of how much the market changed its prices, their demands remain unmet, resulting in low trust towards producers. Additionally, the loan levels remain high. This case is shown in Figure 11.

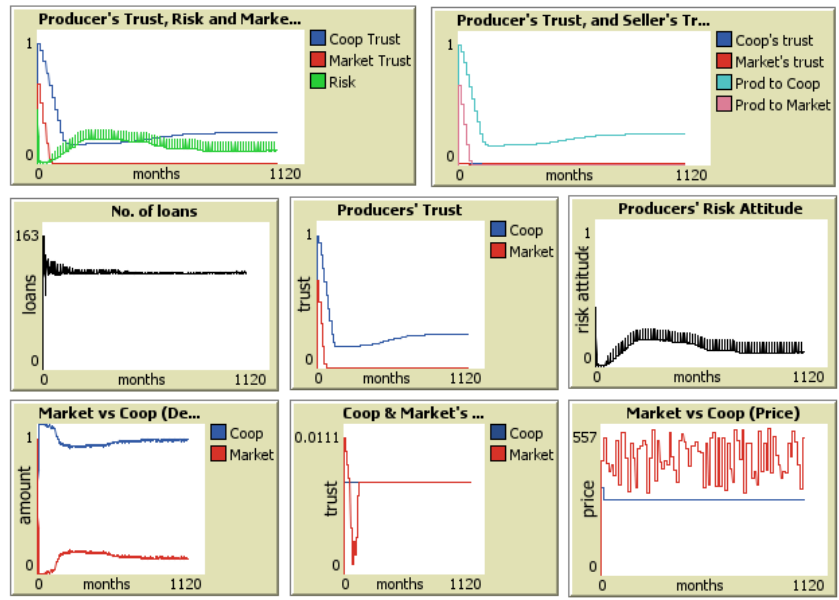

Fig. 11. Case 5, the producer prioritizes their risk-attitude by 100% in NetLogo.

**50% trust, 50% transactional costs.** The simulations show that a higher risk attitude aligns with lower loans but increased price fluctuations. Conversely, a lower risk attitude results in higher loans and, although still present, reduced fluctuations. This indicates that their demands faced challenges in being met, and the buyers experienced more competition.

**50% trust, 50% risk-attitude.** The simulations show that the producer consistently directed sales towards the cooperative, irrespective of market price fluctuations. Consequently, loans typically remained high, accompanied by a prevailing low risk attitude. Despite observing a dip in the cooperative's trust towards producers, most simulations displayed an upward trend.

**50% risk-attitude, 50% transactional costs.** In this scenario, the producer prioritizes their risk-attitude by 50% and buyers' price by 50%. This scenario stood out from others as the buyers' met demands fluctuated within some range, often without overlapping. Nevertheless, the majority of products were consistently sold to the cooperative, resulting in lower prices. Additionally, when the risk attitude was high, the loans were slightly higher. This unique case is shown in figure 12.

*Corresponding Author: groguis@up.edu.ph

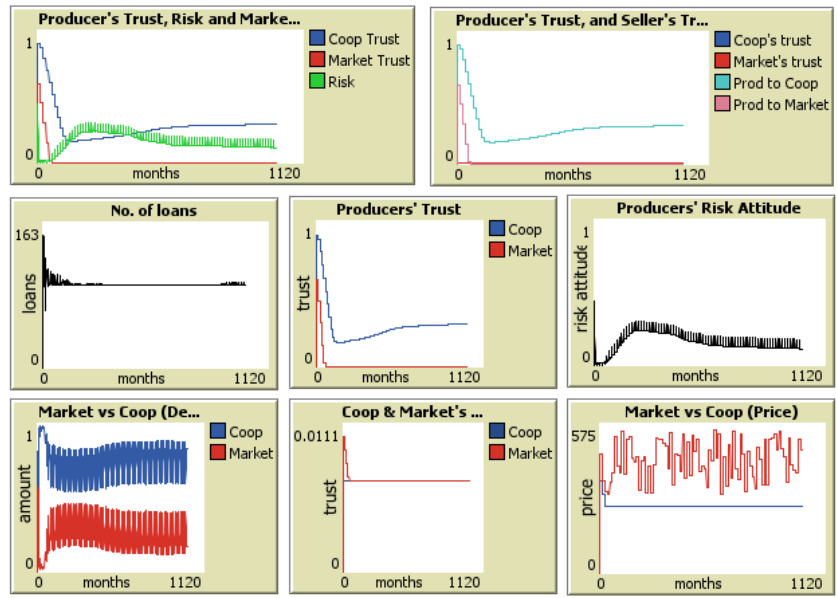

Fig. 12. Case 5, the producer prioritizes their risk-attitude by 50% and the buyer's prize by 50% in NetLogo.

DISCUSSION

The first and second cases did not yield substantial results, particularly when compared to cases 3 to 5. This outcome can be attributed to the fact that in reality, the cooperative is not the sole buyer in the market, and prices are not determined randomly. Despite the assumed properties and parameters in these cases, they offer valuable insights into how trust, risk, and transaction costs influence the value chain.

Regarding trust, producers tend to prefer selling to the cooperative if they place a higher value on trust. While there may be various other factors influencing this decision in reality, a significant aspect is the cooperative's ability to provide loans to producers. Given that capital is essential for coffee production, producers often opt to sell to the cooperative for this reason, even though it may affect their loan repayment capability. Furthermore, an intriguing correlation exists between buyers' demand and their trust in producers, although pinpointing the exact cause proves challenging due to its inconsistency across simulations. Moreover, across all results, compared to risk attitude, trust appears to exert a more significant influence on producers' decisions.

The producers' risk attitude emerges as an intriguing factor within the system. Across all observed outcomes, a notable correlation is evident between the number of loans within the system and the average risk-attitude of the producers. This relationship likely stems from the significant impact of their risk attitude on their ability to engage with alternative markets offering potentially higher prices, thereby affecting their loan repayment capacity. However, if producers successfully meet the cooperative's demand, their risk-attitude may increase, leading to a decrease in sales to the cooperative. This relationship explains the fluctuations in the allocation of supply towards the buyers. This line of reasoning is supported by the observation that as the risk-attitude stabilizes, fluctuations in producers' supply to buyers diminish. When compared to costs, producers prioritize risk-attitude. Additionally, an unexpected yet consistent finding

*Corresponding Author: groguis@up.edu.ph

across all outcomes is the initial rise followed by a subsequent decrease in risk-attitude values.

Lastly, unlike the other two factors, most values related to transaction costs remain constant. However, significant observations emerge from changes in coffee prices and producers' initial funds. Producers with higher off-farm and non-farm incomes are less inclined to seek loans from the cooperative, a logical behavior given the choice. However, it's challenging to determine if these initial values also influence producers' risk attitudes in the same environment. Nevertheless, across all factors, coffee prices notably impact the system. Also, when producers struggle to meet buyer demands, price fluctuations increase, demonstrating market competition. Also, in general, most cases result in a 30-70 allocation of supply. Though, this may be due to the drying capacity of the producers that was noted from the survey.

CONCLUSION

The study aimed to understand the interactions and dynamics among key players in the Coffee Value Chain of Davao del Sur, focusing on trust, risk, and transaction costs. Using an ABM model, it analyzed these factors to gain insights. Data from a pre-test survey questionnaire and secondary were used, although these sources mainly offered perspectives from the producers. Simulations were conducted across five cases and various scenarios, with results systematically compiled. However, due to limited data points, the study can only serve as a proof of concept and lay the groundwork for future ABM research in the value chain.

*Corresponding Author: groguis@up.edu.ph

The stakeholders' decisions about which key factor to prioritize in improving their relationships with others in the value chain will depend on their priorities. For producers, concentrating on prices offered by the buyer and transaction costs can improve loan repayment abilities and reduce debt. Conversely, coffee buyers can achieve favorable outcomes by emphasizing trust and relying on producers' risk attitudes. While fair pricing remains crucial, building stronger, long-term relationships with producers can lead to cost savings for buyers while also meeting their demands. Conversely, loan providers may prioritize fostering mutual trust and assessing the risk attitudes of borrowers. Nevertheless, the simulations demonstrate that it is possible to establish a harmonious system that benefits all parties. However, achieving this requires adjustments to demand, pricing, trust, and risk attitudes of key players, which, in reality, may not be ideal for some parties.

## Recommendations

To improve the study, several recommendations are suggested. Firstly, it's crucial to refine the proxy measures for trust, risk, and transaction costs in the model. This involves enhancing current measures and possibly adding more factors to better grasp these concepts. Gaining more insight into how these factors affect key players would greatly enhance this process. Hence, interviews and key informats may prove more effective compared to survey questionnaires. Additionally, it is necessary to increase the number of data points. The current lack of data points leads to assumptions about aspects such as market properties and price changes. So, expanding the dataset would provide a more accurate understanding of the dynamics at play. Furthermore, while the study primarily focuses on producers' perspectives, it would greatly benefit from capturing the active roles and influences of other key players in the value chain. In the model, the cooperative acts as the loan provider, processor and a market. The one-dimensional approach limits the understanding of how these players are affected. However, the study has demonstrated that these three factors significantly impact the decision-making process of the key players involved, albeit theoretically. Implementing these recommendations will yield more accurate results and contribute to a systematic understanding of the roles of trust, risk, and transaction costs in the coffee value chain.

*Corresponding Author: groguis@up.edu.ph